\documentclass[english,aps,prl,twocolumn,superscriptaddress]{revtex4-1}

\usepackage{bm}
\usepackage{appendix}
\usepackage{graphicx}
\usepackage{amsmath}
\usepackage{amssymb}%
\usepackage{color}
\usepackage{lipsum}
\usepackage[colorlinks=true,citecolor=blue,urlcolor=blue]{hyperref}
\def\cp#1{\mathbf{#1}}

\begin{document}
\title{Generating Giant Vortex in a Fermi Superfluid via Spin-Orbital-Angular-Momentum Coupling}

\author{Ke-Ji Chen}
\thanks{These authors contributed equally to this work.}
\affiliation{Department of Physics and State Key Laboratory of Low-Dimensional
Quantum Physics, Tsinghua University, Beijing 100084, China}

\author{Fan Wu}
\thanks{These authors contributed equally to this work.}
\affiliation{Department of Physics and State Key Laboratory of Low-Dimensional
Quantum Physics, Tsinghua University, Beijing 100084, China}

\author{Shi-Guo Peng}
\affiliation{State Key Laboratory of Magnetic Resonance and Atomic and Molecular Physics, Innovation Academy for Precision Measurement Science and Technology, Chinese Academy of Sciences, Wuhan 430071, China}

\author{Wei Yi}
\email{wyiz@ustc.edu.cn}
\affiliation{CAS Key Laboratory of Quantum Information, University of Science and Technology of China, Hefei 230026, China}
\affiliation{CAS Center For Excellence in Quantum Information and Quantum Physics, Hefei 230026, China}

\author{Lianyi He}
\email{lianyi@mail.tsinghua.edu.cn}
\affiliation{Department of Physics and State Key Laboratory of Low-Dimensional
Quantum Physics, Tsinghua University, Beijing 100084, China}

\begin{abstract}
Spin-orbital-angular-momentum (SOAM) coupling has been realized in recent experiments of Bose-Einstein condensates [Chen \emph{et al}., Phys. Rev. Lett. {\bf 121}, 113204 (2018) and Zhang \emph{et al.}, Phys. Rev. Lett. {\bf 122}, 110402 (2019)], where the orbital angular momentum imprinted upon bosons leads to quantized vortices. For fermions, such an exotic synthetic gauge field can provide fertile ground for
fascinating pairing schemes and rich superfluid phases, which are yet to be explored.
Here we demonstrate how SOAM coupling stabilizes vortices in Fermi superfluids through a unique mechanism that can be viewed as the angular analog to that of the spin-orbit-coupling-induced Fulde-Ferrell state under a Fermi surface deformation. Remarkably, the vortex size is comparable with the beam waist of Raman lasers generating the SOAM coupling, which is typically much larger than previously observed vortices in Fermi superfluids. With tunable size and core structure, these giant vortex states provide unprecedented experimental access to topological defects in Fermi superfluids.
\end{abstract}

\maketitle

In the past decade, the experimental implementation of synthetic spin-orbit coupling (SOC) in cold atoms has stimulated extensive activities in simulating exotic quantum matter~\cite{Lin-11,Zhang-12,Zwierlein-12,socreview1,socreview2,socreview3,socreview4,socreview5,socreview6}. Under a Raman-induced SOC, for instance, the internal hyperfine spins of an atom are coupled with the atomic center-of-mass momentum through a two-photon Raman process, such that the atom is subject to a non-Abelian synthetic gauge field that qualitatively modifies the single-particle dispersion, with nontrivial few- and many-body consequences~\cite{puhantwobody,shenoy,Wu-13,chuanwei-13,Wei-13,tfflo0,tfflohu,soc3body1,soc3body2}.
Recently, an alternative type of synthetic gauge field is introduced where hyperfine spins are coupled with the atomic center-of-mass angular momentum~\cite{Hu-15, Pu-15, Qu-15, Sun-15, Chen-16, Hu-19, Chen-19}. Such a spin-orbital-angular-momentum (SOAM) coupling has already been experimentally demonstrated in spinor Bose-Einstein condensates (BECs), under a Raman process driven by two Laguerre-Gaussian lasers with different orbital angular momenta~\cite{Lin-18, Jiang-19}. While it has been shown that the SOAM coupling leads to spin-dependent vortex formation and rich quantum phases in BECs~\cite{Hu-15, Pu-15, Qu-15, Sun-15, Chen-16, Hu-19, Chen-19, Han-20, Lin-18, Jiang-19}, its impact on fermions is yet to be explored. Particularly, vortices in a SOAM coupled Fermi gas would provide crucial insight to the underlying superfluid.
However, unlike the SOAM-coupled BEC, SOAM coupling alone does not induce vortices in a Fermi superfluid, since fermions in a Cooper pair would acquire opposite orbital-angular momenta that cancel each other, yielding a superfluid devoid of vortices.

In this Letter, we demonstrate a unique vortex-forming scheme for a Fermi superfluid under SOAM coupling.
We show that stable vortices {\it can} be generated under experimentally achievable parameters by introducing a moderate two-photon detuning in the Raman-induced SOAM coupling.
The underlying mechanism of the vortex formation is reminiscent of that of the SOC-induced Fulde-Ferrell pairing, where Cooper pairs inevitably carry finite center-of-mass momentum due to the asymmetry of the SOC-dressed Fermi surface under Zeeman fields~\cite{puhantwobody,shenoy,Wu-13,chuanwei-13,Wei-13,tfflo0,tfflohu}. An angular analog of the Fulde-Ferrell pairing state, SOAM-coupling-induced vortices feature Cooper pairs with quantized angular momenta, and are easier to detect due to their topological nature.
Compared to previously known cases in Fermi superfluids~\cite{type2-vortex,Ketterle-vortex}, vortices here are induced by non-Abelian rather than Abelian gauge fields.
The pairing mechanism here is also distinct from that of Fulde-Ferrell-Larkin-Ovchinnikov (FFLO)  states in a polarized superfluid, as FFLO states do not rely on Fermi-surface asymmetry and are much less stable~\cite{FF,LO,socreview4}.

As a consequence of this exotic vortex-forming process, the SOAM-coupling-induced vortices possess fascinating features.
First, these vortices are strikingly large in size, with the order parameter or density profiles varying on a length scale comparable to the waist of Raman beams. This is markedly different from previously studied vortices in atomic Fermi superfluids, where changes in the vortex-core structure predominantly take place within a short
length scale set by the interatomic separation~\cite{Ho-vortex,Levin-vortex}.
Second, the vortex core exhibits a large spin imbalance, which originates from spin-polarized vortex bound states, or the so-called Caroli--de Gennes--Matricon (CdGM) states~\cite{Caroli-64}, and would serve as an ideal experimental signal.
Our results offer an intriguing way of generating vortices with tunable size in a Fermi superfluid,
and would stimulate further study of exotic pairing states in a SOAM coupled Fermi gas.

\begin{figure}[t]
\begin{center}
\includegraphics[width=0.48\textwidth]{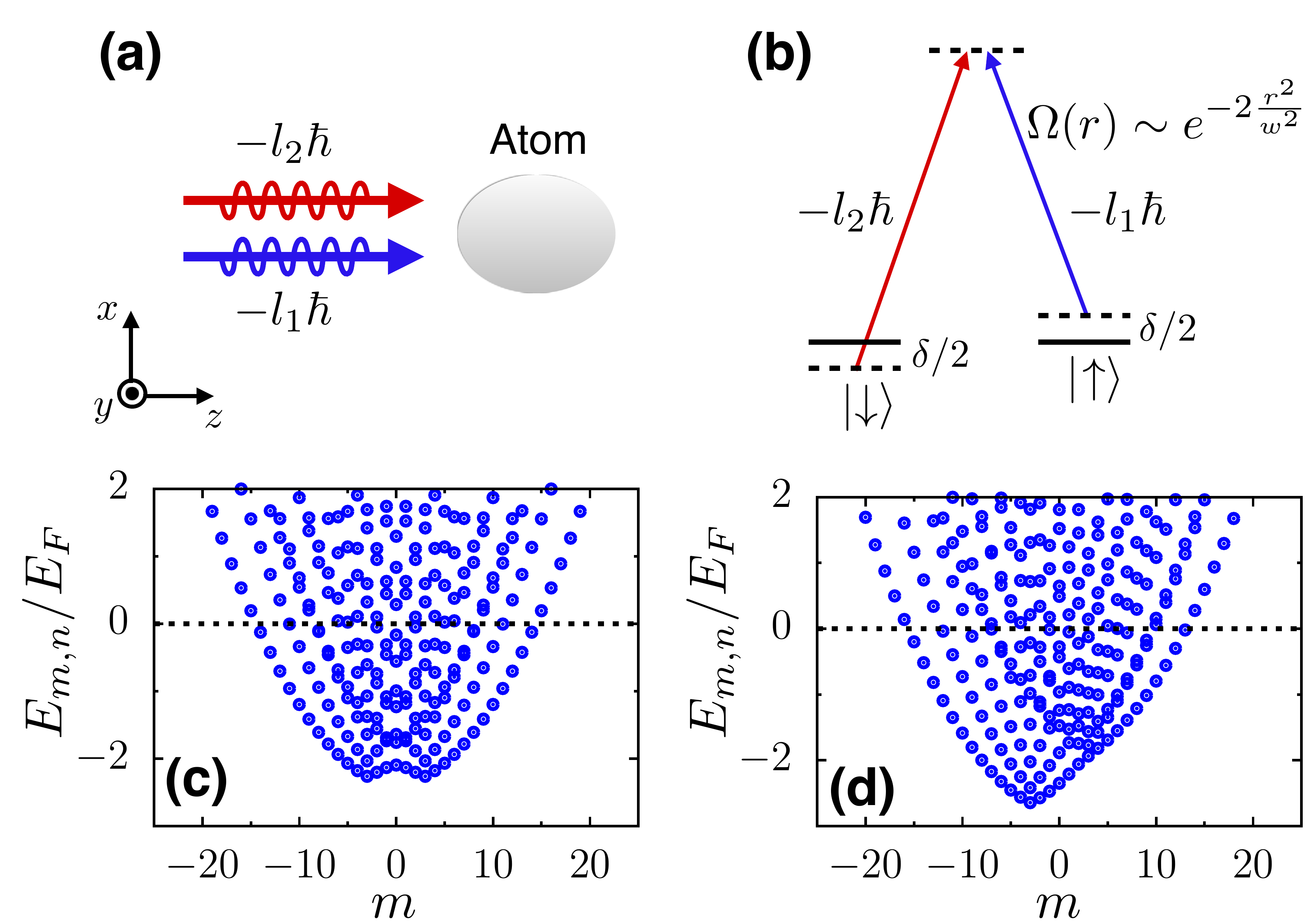}
\caption{(a) A pair of copropagating Raman beams carrying different orbital angular momenta ($-l_1\hbar$ and $-l_2\hbar$) induce SOAM coupling in atoms, with a transferred angular momentum $2l\hbar=(l_1-l_2)\hbar$. (b) Schematic illustration of the level scheme.
(c)(d) Single-particle energy spectra under SOAM coupling for $\delta=0$ (c), and $\delta /E_F= 0.4$ (d). Black dashed lines denote potential Fermi surfaces in a many-body setting.
The parameters are $l=3$, $\Omega_0/E_F=0.2$,  $k_F w = 15$,  and $k_F R=15$.
Here the Fermi wave vector $k_F$ is defined through the density $n_0=k_F^2/(2\pi)$ of a uniform Fermi gas in the area $S=\pi R^2$, and $E_F=\hbar^2k_F^2/(2M)$.}
\label{Fig1}
\end{center}
\end{figure}


{\it Model.---}
We consider a two-component Fermi gas confined in the two-dimensional $x$-$y$ plane. The SOAM coupling is driven by a pair of copropagating Raman beams carrying different orbital angular momenta $-l_1\hbar$ and $-l_2\hbar$ [see Figs.~\ref{Fig1}(a),\ref{Fig1}(b)]. The two-photon Raman process is characterized by an inhomogeneous Raman coupling $\Omega(r)$ and a phase winding $e^{-2il\theta}$, where $2l\equiv l_1-l_2$ and we adopt the polar coordinates ${\bf r}=(r,\theta)$. After a unitary transformation, the effective single-particle Hamiltonian becomes~\cite{Jiang-19,supp}
\begin{align}
{\cal H}_s=&-\frac{\hbar^2}{2Mr}\frac{\partial}{\partial r}\left(r\frac{\partial}{\partial r}\right)+\frac{(L_z-l\hbar\sigma_z)^2}{2Mr^2}\nonumber\\
&+\ \Omega(r)\sigma_x+\frac{\delta}{2} \sigma_z+V_{\text{ext}}({\bf r}),
\label{Hs}
\end{align}
where $M$ is the atom mass, and $\sigma_i$ ($i=x,y,z$) are the Pauli matrices. The atomic orbital angular momentum perpendicular to the $x$-$y$ plane, $L_{z}=-i\hbar \partial /\partial \theta$, is coupled to the atomic spin through the SOAM-coupling term $-l\hbar L_z\sigma_z/(Mr^2)$. The atoms are subject to an external potential $V_{\rm ext}({\bf r})$,  which we take as an isotropic hard-wall box potential with a radius $R$.

The Raman coupling $\Omega(r)$ and two-photon detuning $\delta$
provide effective transverse and longitudinal Zeeman fields, respectively, which play key roles in stabilizing vortices. While the Laguerre-Gaussian Raman beams were used in previous experiments~\cite{Lin-18, Jiang-19}, their intensities are suppressed over a considerable region near $r=0$, leading to vanishingly small SOAM-coupling effects in the vicinity which we find unfavorable for vortex formation. Instead, we propose to directly image Raman beams with distinct orbital angular momenta onto atoms, before the beams propagate into the diffraction far field~\cite{supp}. The resulting Raman coupling $\Omega(r)$ has an overall Gaussian profile, but with a small core of suppressed intensity whose size can be 2 orders of magnitude smaller than the laser waist~\cite{Rumala-17}.
It is then a good approximation to write the Raman coupling as $\Omega(r)=\Omega_0 e^{-2r^2/w^2}$, with $\Omega_0$ the peak intensity and $w$ the beam waist~\cite{supp}.
Assuming Raman lasers operate at the tune-out wavelength~\cite{tuneout1,tuneout2,Jiang-19}, we neglect ac Stark shifts.

Considering $s$-wave contact interactions between different spin components, we write the many-body Hamiltonian as $H=H_0+H_{\text{int}}$, with $H_0=\int d{\bf r} \Psi^{\dag}(\cp r) {\cal{H}}_s \Psi(\cp r)$, and $H_{\text{int}}=-g\int d{\bf r}\psi^{\dag}_{\uparrow}({\bf r})\psi^{\dag}_{\downarrow}({\bf r})\psi_{\downarrow}({\bf r})\psi_{\uparrow}({\bf r})$. Here
$\Psi(\cp r)=[\psi_\uparrow(\cp r),\ \psi_\downarrow(\cp r)]^{\rm T}$, with $\psi_{\sigma}(\cp r)$ ($\sigma=\uparrow,\downarrow$) denoting the field operators for the two hyperfine spins.
The bare interaction $g$ is renormalized as $g = 4 \pi \hbar^2 /[M \ln (1+2 E_{c}/E_{B} )]$~\cite{supp,Randeria1989,Randeria1990,2drenorm,Cuirenorm}, where $E_{B}$ is the two-body binding energy in the absence of SOAM coupling, and $E_c$ is a large energy cutoff that does not affect our results.

{\it Single-particle spectrum.---}
Before moving to the many-body calculation, we first analyze the single-particle properties, i.e., the spectrum of ${\cal H}_s$, which is helpful for understanding the mechanism of vortex formation. Because of  the rotational symmetry, the eigen wave function of ${\cal H}_s$ can be written as $\psi_{mn}({\bf r})=\varphi_{mn}(r)\Theta_{m}(\theta)$, where the angular and radial wave functions are $\Theta_{m}(\theta)=e^{im\theta}/\sqrt{2\pi}$ and $\varphi_{mn}(r)$, respectively, with quantum numbers $m\in\mathbb{Z}$ and $n\in \mathbb{Z}^+$.
We numerically solve the Schr\"odinger equation ${\cal H}_s \psi_{mn}({\bf r})=E_{m,n} \psi_{mn}({\bf r})$ in the basis of the Bessel functions to determine the energy spectrum $E_{m,n}$. This amounts to writing
$\varphi_{mn}(r)=[f_{\uparrow mn}(r), f_{\downarrow mn}(r)]^{\rm T}$, and making the expansion $f_{\sigma mn}(r) =  \sum_{n'}c^{(n')}_{\sigma mn}R_{n',m-l\tau}(r)$, where $R_{n,m}(r) =  \sqrt{2}J_{m}(\alpha_{nm}r/R)/R J_{m+1}(\alpha_{nm})$, with $\tau=+1\ (-1)$ for $\sigma=\uparrow(\downarrow)$. Here $J_m(x)$ is the Bessel function of the first kind whose zeros are given by $\alpha_{nm}$.

In Figs.~\ref{Fig1}(c),\ref{Fig1}(d), we show the impact of $\delta$. For $\delta=0$, an inversion symmetry exists in the eigenspectrum, leading to a symmetric distribution of eigenenergies with respect to $m=0$,
with $E_{m,n}=E_{-m,n}$. In contrast, for nonzero $\delta$, the inversion symmetry is broken, with $E_{m,n}\neq E_{-m,n}$ and hence a deformed Fermi surface in the many-body setting. For either a small or very large $l$ under fixed Zeeman fields, the Fermi-surface deformation is
not significant enough to have an appreciable impact on vortex formation~\cite{supp}. We therefore take an intermediate $l=3$ throughout the work.

When the attractive interaction is turned on, pairing should predominantly take place between unlike spins with the same radial quantum number $n$ to maximize the overlap of radial wave functions.
Thus, for the symmetric eigenspectrum under $\delta=0$, it is more favorable for two fermions with opposite angular quantum numbers ($m$ and $-m$) to form a Cooper pair carrying a vanishing total angular momentum.
In contrast, under a finite $\delta$ with asymmetric eigenspectrum, the two fermions in a Cooper pair may possess different values of $|m|$, resulting in a pairing state with a nonzero, quantized angular momentum, which is nothing but a vortex state.
Such a mechanism for the vortex formation is analogous to that of the SOC-induced Fulde-Ferrell states, where the interplay between SOC and Zeeman fields leads to the deformation of Fermi surfaces with broken inversion symmetry in the momentum space~\cite{Wu-13,chuanwei-13,Wei-13,tfflo0,tfflohu}. However, a key difference here is the quantization of the angular momentum, which gives rise to topological defects in the resulting Fermi superfluid.

{\it Many-body formalism.---}
We confirm the analysis above by solving the many-body problem under the Bogoliubov--de Gennes (BdG) formalism. The BdG equation is given by ${\cal H}_{\text{BdG}}\Phi_{mn}({\bf r})=\epsilon_{mn}\Phi_{mn}({\bf r})$, with
\begin{align}
{\cal H}_{\text{BdG}}
=\left[\begin{array}{cccc}K_{\uparrow }({\bf r}) &\Omega(r) & 0 & \Delta({\bf r}) \\\Omega(r) & K_{\downarrow }({\bf r}) & -\Delta({\bf r}) & 0 \\0 & -\Delta^{\ast}({\bf r}) & -K^{\ast}_{\uparrow}({\bf r}) & -\Omega(r) \\ \Delta^{\ast}({\bf r}) & 0 & -\Omega(r) & -K^{\ast}_{\downarrow}({\bf r})\end{array}\right].
\label{HC}
\end{align}
Here $\Phi_{mn}({\bf r})=[ u_{\uparrow mn}, u_{\downarrow mn} , v_{\uparrow mn} , v_{\downarrow mn}]^{\rm T}$, with $u_{\sigma mn}$ and $v_{\sigma mn}$ being the Bogoliubov coefficients, and
$K_{\sigma}({\bf r})  =  -\frac{\hbar^2}{2M}\left[\frac{1}{r}\frac{\partial}{\partial r}(r\frac{\partial}{\partial r})+\frac{1}{r^2}\left(\frac{\partial}{\partial \theta}-i\tau l \right)^2\right]+\tau \frac{\delta}{2}-\mu$. The chemical potential $\mu$ is
introduced to fix the total particle number $N$. The superfluid order parameter $\Delta({\bf r})=-g \langle \psi_{\downarrow}({\bf r}) \psi_{\uparrow}({\bf r})\rangle$ is determined self-consistently,
\begin{align}
\Delta({\bf r})   =  \frac{g}{2}\sum_{mn}\Big[u_{\uparrow mn}v^* _{\downarrow mn}  \vartheta (\epsilon_{mn})+u_{\downarrow mn} v^*_{\uparrow mn }\vartheta(-\epsilon_{mn})\Big],
\label{eq:delta}
\end{align}
where $\vartheta (x)$ is the Heaviside step function.

\begin{figure}
\begin{center}
\includegraphics[width=0.4\textwidth]{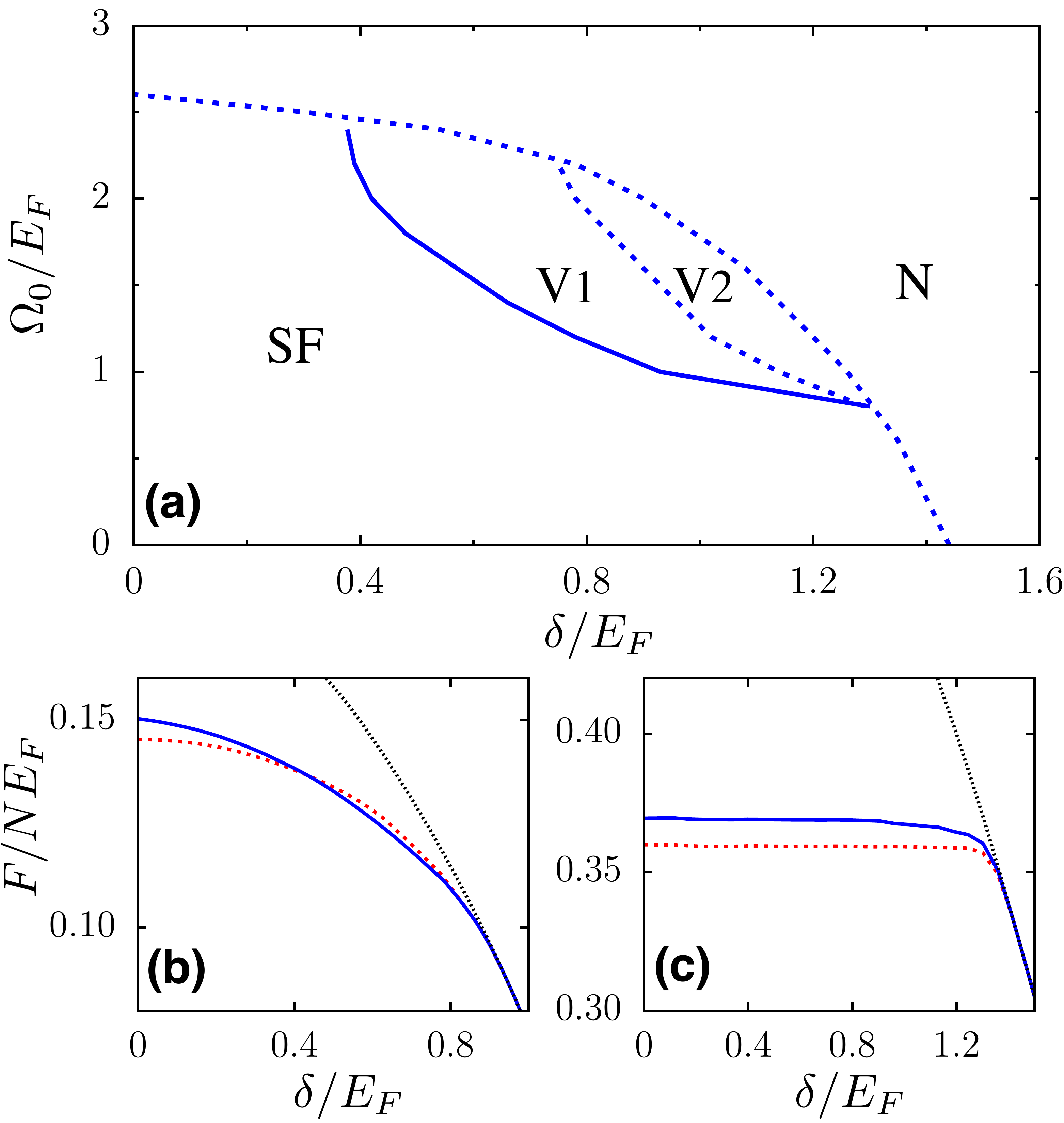}
\caption{(a) Phase diagram of a two-dimensional Fermi superfluid with SOAM coupling in the $\Omega_0$-$\delta$ plane.We fix $l=3$, and the interaction strength is chosen as $E_B/E_F=0.5$. The phase diagram includes the usual superfluid state  with $\kappa=0$, the normal state  with $\Delta=0$, and two vortex states with $\kappa=-1$: a fully gapped vortex states ($V1$) and a gapless vortex state ($V2$).
(b)(c) Free energies $F$ of the superfluid ($\kappa=0$; red dashed), vortex ($\kappa=-1$; blue solid) and normal (black dotted) states as functions of $\delta$, with $\Omega_0/E_F=2$ (b) and $\Omega_0/E_F=0.5$ (c).
The parameters $k_Fw$ and $k_FR$ are the same as those in Fig.~\ref{Fig1}.}
\label{Fig2}
\end{center}
\end{figure}

\begin{figure}[t]
\begin{center}
\includegraphics[width=0.48\textwidth]{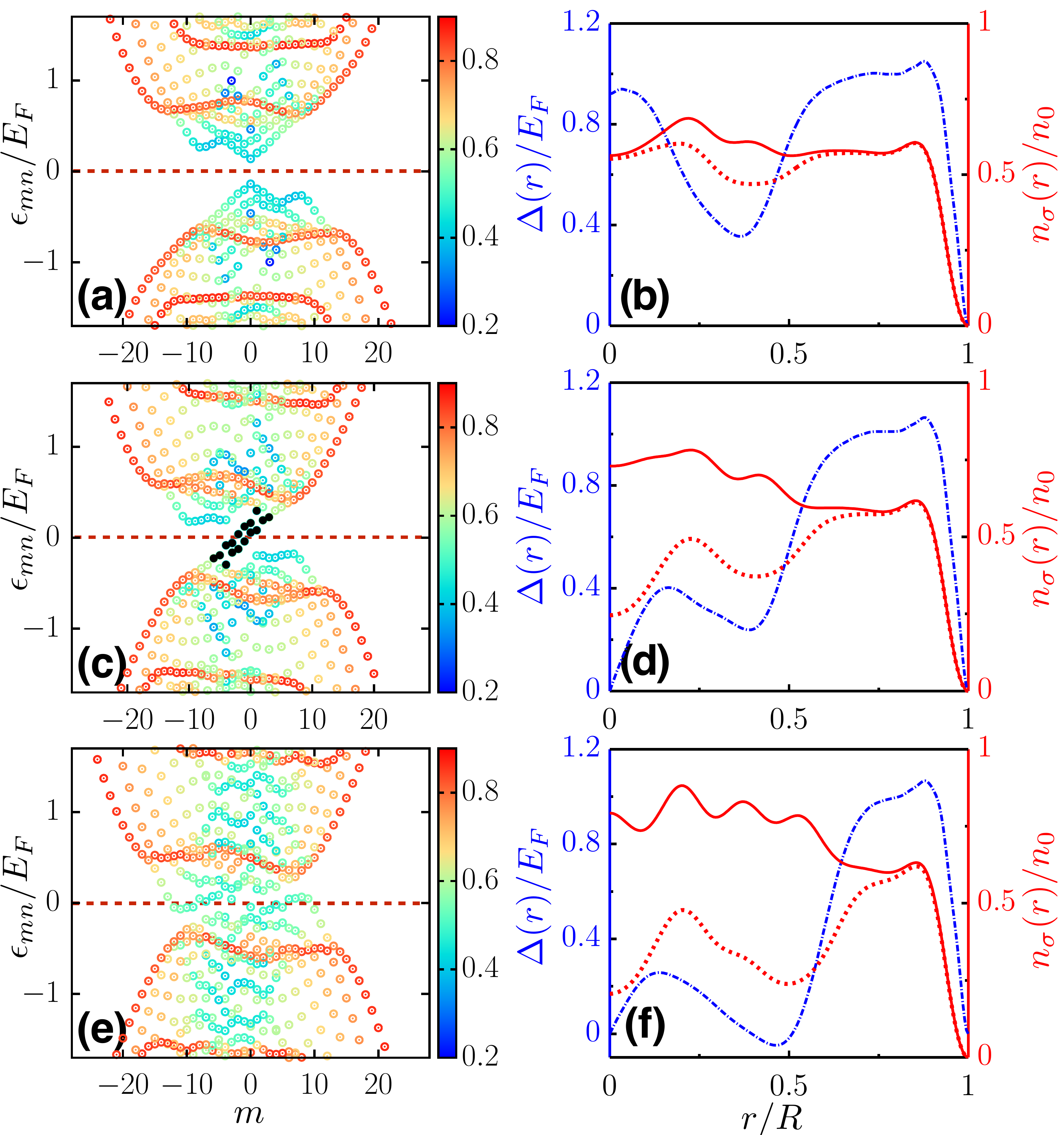}
\caption{Energy spectra and vortex-core structures for the SF state with  $\delta/E_F=0.48$ [(a),(b)],  the $V1$ state with $\delta/E_F=0.84$ [(c),(d)], and the $V2$ state with $\delta/E_F=1.05$ [(e),(f)]. In the energy spectra (a),(c),(e), different eigenstates are color coded according to the ratio $\sqrt{\langle r^2 \rangle}/R$, with the root-mean-square of the radius
$\langle r^2 \rangle =\sum_{\sigma} \int d {\bf r} r^2 (|u_{\sigma mn}|^2+|v_{\sigma mn}|^2)$.
The CdGM states are indicated by black dots in (c).
In (b),(d),(f), the blue dash-dotted correspond to the order parameter profile, and red solid (dashed) lines
are the radial density profiles of the spin-down (up) atoms, respectively. The radial density is calculated through $n_{\sigma}(r)=\frac{1}{2\pi} \int d\theta n_{\sigma}({\bf r}) $. Here we take $\Omega_0/E_F=1.2$ and other parameters are the same as those in Fig.~\ref{Fig2}.}
\label{Fig3}
\end{center}
\end{figure}

\begin{figure}[tbp]
\begin{center}
\includegraphics[width=0.4\textwidth]{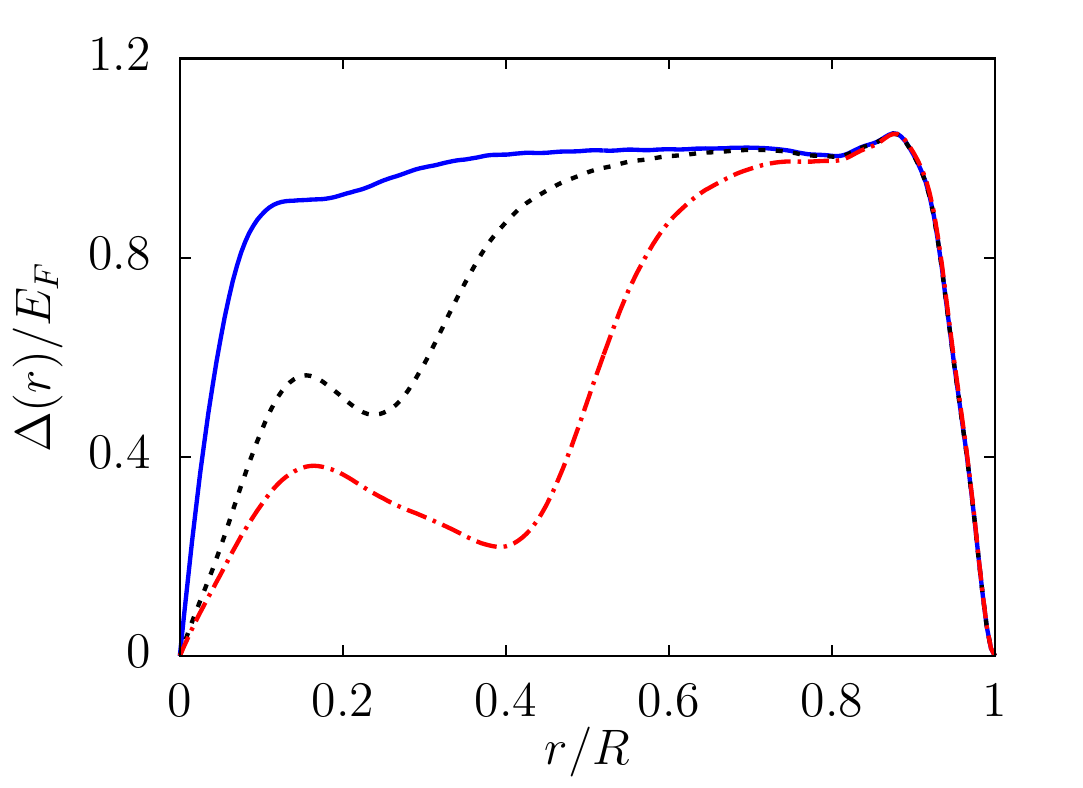}
\caption{Order parameter profiles for $k_Fw=5$ (blue solid), $k_Fw=10$ (black dashed), and $k_Fw=15$ (red dash-dotted). We fix $\delta/E_F=0.84$ and $k_FR=15$, with other parameters the same as those in Fig.~\ref{Fig3}.}
\label{Figvortex}
\end{center}
\end{figure}

For a vortex state, we have
$\Delta({\bf r})=\Delta(r)e^{i \kappa \theta}$, where the vorticity $\kappa \in \mathbb{Z}$ characterizes the quantized angular momentum and serves as the topological invariant for the vortex state. We then write $u_{\sigma mn}  = \sum_{ n'}c^{(n')}_{\sigma mn}R_{n', m-\tau l }(r)\Theta_{m}(\theta)$,
and $v_{\sigma mn }=\sum_{ n'}d^{(n')}_{\sigma mn}R_{n',m+\tau l -\kappa}(r)\Theta_{m-\kappa}(\theta)$, so that the BdG equation becomes a matrix equation for the coefficients $c^{(n')}_{\sigma mn}$ and $d^{(n')}_{\sigma mn}$ ~\cite{supp}.

With different values of $\kappa \in \mathbb{Z}$, we solve the BdG equation and the self-consistent equation (\ref{eq:delta}) under the particle number constraint $N=\sum_\sigma\int d\cp r n_\sigma(\cp r)$, with density profiles $n_{\sigma}({\bf r})  =  \frac{1}{2}\sum_{mn}[ |u_{\sigma mn}|^2\vartheta(-\epsilon_{mn})+|v_{\sigma mn}|^2 \vartheta(\epsilon_{mn})]$. We then compare the free energies of vortex states with different $\kappa$, the usual superfluid state ($\kappa=0$), and the normal state ($\Delta=0$) to determine the ground state.

{\it Phase diagram and vortex structure.---}
In Fig.~\ref{Fig2}(a), we show typical phase diagram of the system in the $\Omega_0$--$\delta$ plane. For $\delta>0$ ($\delta<0$),
vortex states with $\kappa=-1$ ($\kappa=1$) are favored, with phase boundaries unchanged by the sign of $\delta$. The fact $|\kappa|\neq l$ indicates that the vortex state here is a many-body phenomenon.
At small $\Omega_0$ and $\delta$, the ground state is a usual superfluid (SF) with a vanishing vorticity $\kappa=0$.
Under sufficiently large $\Omega_0$ and/or $\delta$, the free-energy difference between the SF and normal ($N$) states becomes vanishingly small. Since beyond-mean-field fluctuations tend to stabilize the normal state, for all practical purposes, we consider the system to be in a normal state when the free-energy difference is smaller than $10^{-3}E_F$~\cite{Wu-13}.
More importantly, between SF and $N$ states, two vortex states
exist. For example, with a fixed $\Omega_0/E_F=1.5$ [see Fig.~\ref{Fig2}(a)], the ground state is in the SF state under small detunings $\delta$, and becomes a fully-gapped vortex state ($V1$) beyond a critical value of $\delta$.
Further increase of $\delta$ leads to a gapless vortex state ($V2$), where the bulk excitation gap closes.

In Figs.~\ref{Fig2}(b),\ref{Fig2}(c), we compare free energies of different states, as the phase diagram is traversed.
In particular, for the case with $\Omega_0=0.5E_F$, the ground state remains vortexless for finite $\delta$, despite the deformation of the Fermi surface under SOAM coupling and effective Zeeman fields.
This is due to the quantized nature of the angular momentum, and is in sharp contrast to the SOC-induced Fulde-Ferrell state which necessarily acquires a finite, continuously varying center-of-mass momentum in the presence of SOC and Zeeman fields~\cite{puhantwobody,shenoy,Wu-13,chuanwei-13,Wei-13,tfflo0,tfflohu}.

In Fig.~\ref{Fig3}, we show the energy spectrum $\epsilon_{mn}$, the order parameter $\Delta(r)$, and the density profiles in the SF state [Figs.~\ref{Fig3}(a),\ref{Fig3}(b)], the $V1$ state [Figs.~\ref{Fig3}(c),\ref{Fig3}(d)], and the $V2$ state [Figs.~\ref{Fig3}(e),\ref{Fig3}(f)]. In the ground state, all eigenstates with $\epsilon_{mn}<0$ are occupied. Here two important observations are in order.

First, both the order parameter and density profiles vary over a length scale set by the laser waist $w$,
leading to a giant vortex core with tunable size. This is clearly visible in Fig.~\ref{Figvortex}, where the vortex-core size, characterized by variations of the order parameter, is comparable to $w$, the latter being much larger than $k_F^{-1}$ in experiments.

\begin{figure}[tbp]
\begin{center}
\includegraphics[width=0.48\textwidth]{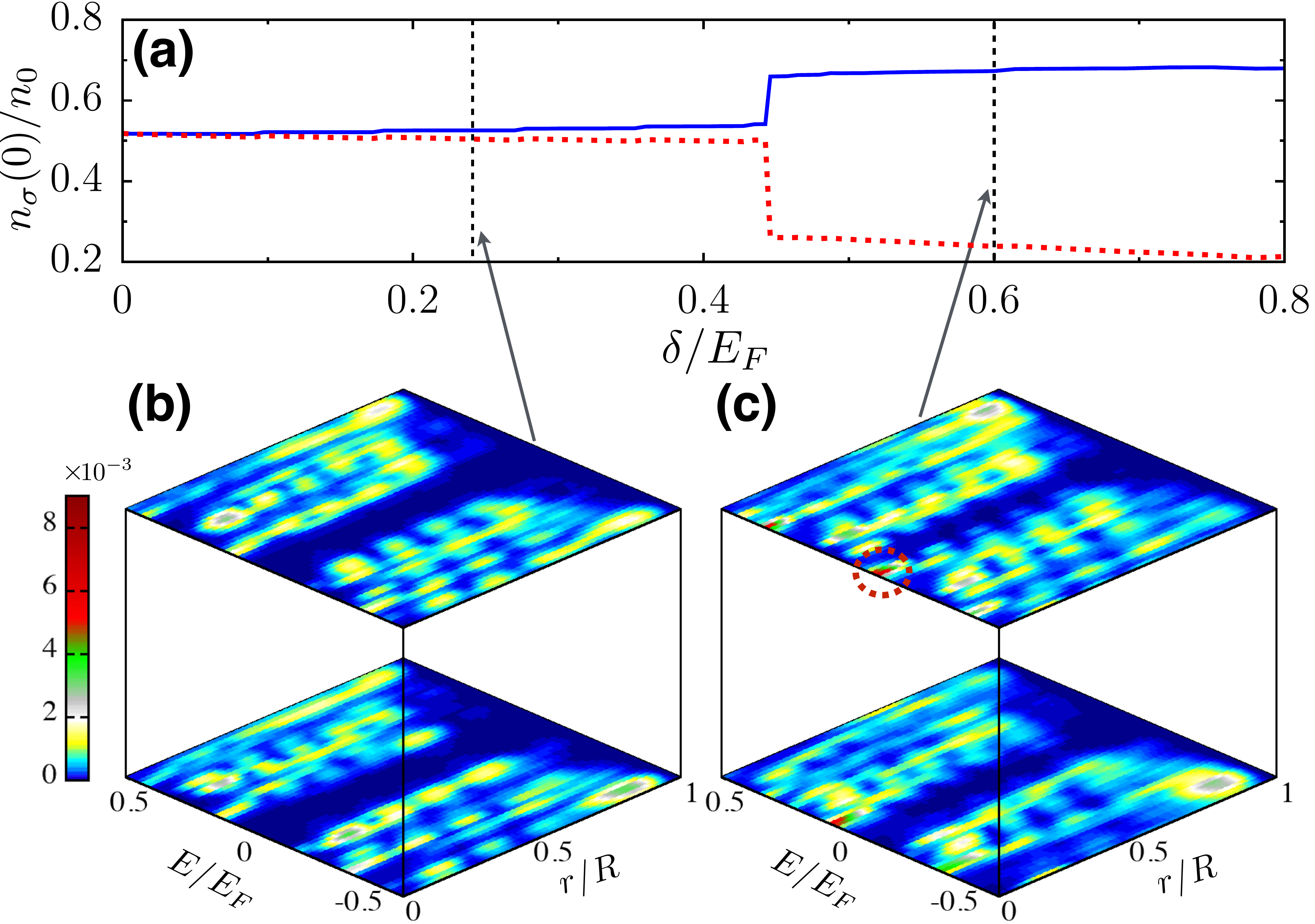}
\caption{(a)Densities of the two spin components at the trap center, $n_{\sigma}(r=0)$, as a function of the detuning $\delta$. The blue solid (red dashed) curve denotes the spin-down (spin-up) component.
(b),(c) Local densities of state $D_{\sigma}(r, E)$ for $\delta/E_F=0.24$ (b) and $\delta/E_F=0.6$ (c). The upper (lower) layer shows the local density of state of the spin-down (spin-up) component. The red circle in (c) indicates the occupied CdGM state responsible for the spin imbalance at $r=0$. Here we take $\Omega_0/E_F=2$ and other parameters are the same as those in Fig.~\ref{Fig2}.}
\label{Fig4}
\end{center}
\end{figure}

Second, the order parameter $\Delta(r)$ of either vortex states vanishes at $r=0$, accompanied by a large spin imbalance near the vortex core. This is in contrast to the SF state, where the spin population is roughly balanced at $r=0$ [Fig.~\ref{Fig3}(b)].
In Fig.~\ref{Fig4}(a), we show the densities of the two spin components at $r=0$ as a function of the detuning $\delta$. The spin imbalance remains small in the SF state at small $\delta$, but abruptly jumps to a large value once the system is in the vortex state, suggesting a discontinuous phase transition. To better characterize the large spin polarization at the vortex core, we calculate the local density of state $D_{\sigma}(r, E) =\frac{1}{2\pi} \int d\theta D_{\sigma}({\bf r},E)$, which offers spatial and spectral resolutions of all eigenstates. Here $D_{\sigma}({\bf r}, E)  =  \frac{1}{2}\sum_{mn}[|u_{\sigma mn}|^2 \delta(E-\epsilon_{mn})+|v_{\sigma mn}|^2 \delta(E+\epsilon_{mn})]$. Our results for the SF and $V1$ states are shown in Figs.~\ref{Fig4}(b) and \ref{Fig4}(c), respectively.
As indicated by the red circle in Fig.~\ref{Fig4}(c), occupied CdGM states with large spin polarization
exist in the excitation gap, which are localized near $r=0$ with depleted pairing order parameter, effectively serving as a dump for the spin polarization under Zeeman fields. These CdGM states are directly responsible for the observed large spin polarization at the vortex core in Fig.~\ref{Fig3}(d). In the $V2$ state, despite the closing of the bulk excitation gap, local spin polarization near the core persists, due to polarized local states that are smoothly connected to the CdGM states in the gapped phase~\cite{supp}.


{\it Discussion.---}
Experimentally, the SOAM-coupling-induced vortices can be detected through the apparent spin polarization at vortex cores, which is further facilitated by their large and tunable size. For instance, with $w\sim 7.5$  $\mu $m and $l=3$, and under typical parameters of a quasi-two-dimensional Fermi gas of $^{6}$Li atoms with  $E_F \sim 2\pi \hbar \times 3.4$kHz~\cite{Fermigas2D}, the parameter window for a stable vortex state is $\Omega_0  \sim 2\pi \hbar \times (3.4, 8.5)$ and
$\delta \sim 2\pi\hbar \times (1.4, 4.1)$kHz, which are readily accessible.
While we focus on the case with $l=3$, vortex states with $|\kappa|>1$ may be stabilized under a larger $l$~\cite{supp},  where one should then consider the possibility of multiple vortices. Our work thus offers the prospect of engineering more exotic vortex states via the SOAM coupling.

{\it Note added.}  Recently, a related preprint appeared~\cite{jeevortex}, where a similar vortex-forming scheme under SOAM coupling is studied in a different configuration.

\begin{acknowledgements} We acknowledge fruitful discussions with Jian-Song Pan, Jing Zhou, Dongyang Yu, and Tian-You Gao. This work is supported by the National Key R\&D Program (Grants No. 2018YFA0306503, No. 2016YFA0301700, No. 2017YFA0304100, No. 2016YFA0301503) and the Natural Science Foundation of China (Grants No. 11775123, No. 11974331, No. 11974384).
\end{acknowledgements}

\bibliographystyle{apsrev4-1}

\widetext

\begin{center}
\textbf{\large  Supplemental Material for  ``Generating Giant Vortex in a Fermi Superfluid via Spin-Orbital-Angular-Momentum coupling"}
\end{center}
\setcounter{equation}{0}
\setcounter{figure}{0}
\setcounter{table}{0}
\makeatletter
\renewcommand{\theequation}{S\arabic{equation}}
\renewcommand{\thefigure}{S\arabic{figure}}
\renewcommand{\citenumfont}[1]{#1}

In this Supplemental Material, we provide details on the experimental implementation, derivation of the effective single-particle Hamiltonian, renormalization of the bare interaction strength, the matrix form of the BdG equation,
the spectrum of the gapless vortex state, and discussions on the vorticity of the vortex state.

\begin{figure}[t]
\includegraphics[width=0.8\columnwidth]{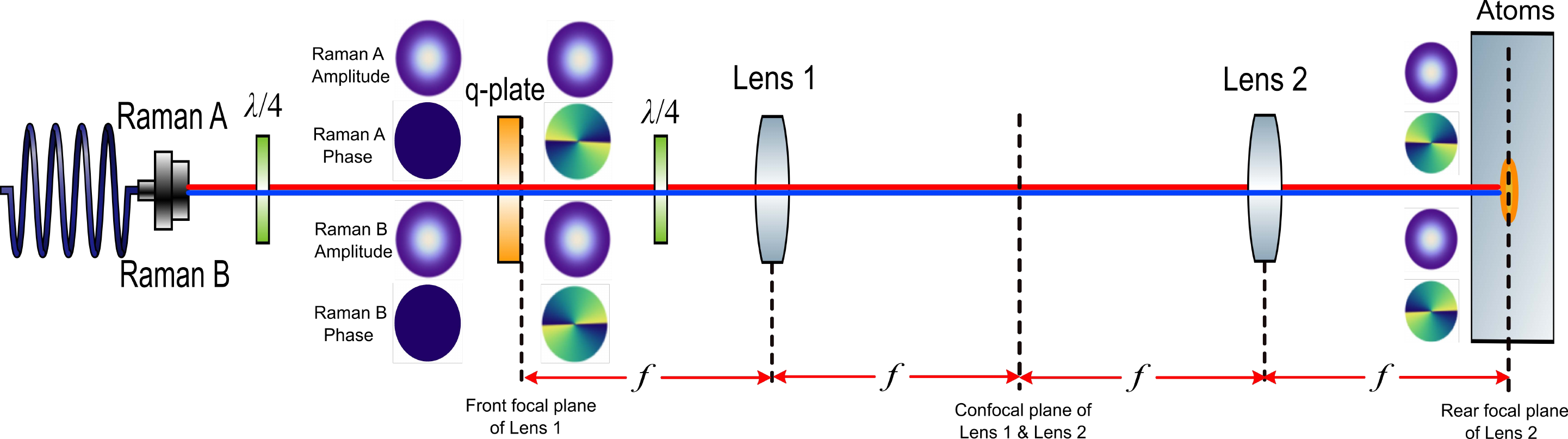}
\caption{A simplified illustration of the proposed optical setup. A pair of
Gaussian Raman beams, Raman $A$ and Raman $B$, emerge from the same
optical fiber with orthogonal linear polarizations (the red and blue
straight lines with slight lateral shift to guide the eye). They are
transformed into circularly polarized ones by a quarter-wave plate
to acquire a winding phase $e^{\pm il\theta}$ from the $q$-plate
(depending on their polarizations). The Raman beams are then transformed
back to linear polarizations again by a quarter-wave plate.
The Gaussian-type profiles right after the $q$-plate with winding
phases $e^{\pm il\theta}$ are imaged by a $4f$-lens system onto the
atoms placed on the rear focal plane.}
\label{fig:ExpSetUp}
\end{figure}

\section{Experimental implementation}

In this section, we provide details on an experimental proposal for implementing a Gaussian-type Raman coupling.
As illustrated in Fig.~\ref{fig:ExpSetUp}, phase patterns $e^{+il\theta}$
and $e^{-il\theta}$ are imprinted by a $q$-plate respectively on
the input pair of Gaussian beams with different circular polarizations. At this stage, the laser beams would still preserve a Gaussian intensity profile, except for a very small region near $r=0$, i.e., at the optical vortex core, where the optical density is attenuated.
A $4f$-lens system is then employed to image the Gaussian intensity profile immediately after the $q$-plate onto the atoms~\cite{Rumala-17}.
Unlike the situations in previous works~\cite{Lin-18, Jiang-19} where laser beams further propagate into the diffraction far field and acquire an intensity profile that can be well-approximated by the Laguerre-Gaussian function,
here the laser beams are imaged onto atoms immediately after passing through the $q$-plate. According to Ref.~\cite{Rumala-17}, the resulting laser beam still possesses an overall Gaussian profile, but with a small central core that can be two orders of magnitude smaller in size than that of the laser waist.

\begin{figure}
\includegraphics[width=0.8\columnwidth]{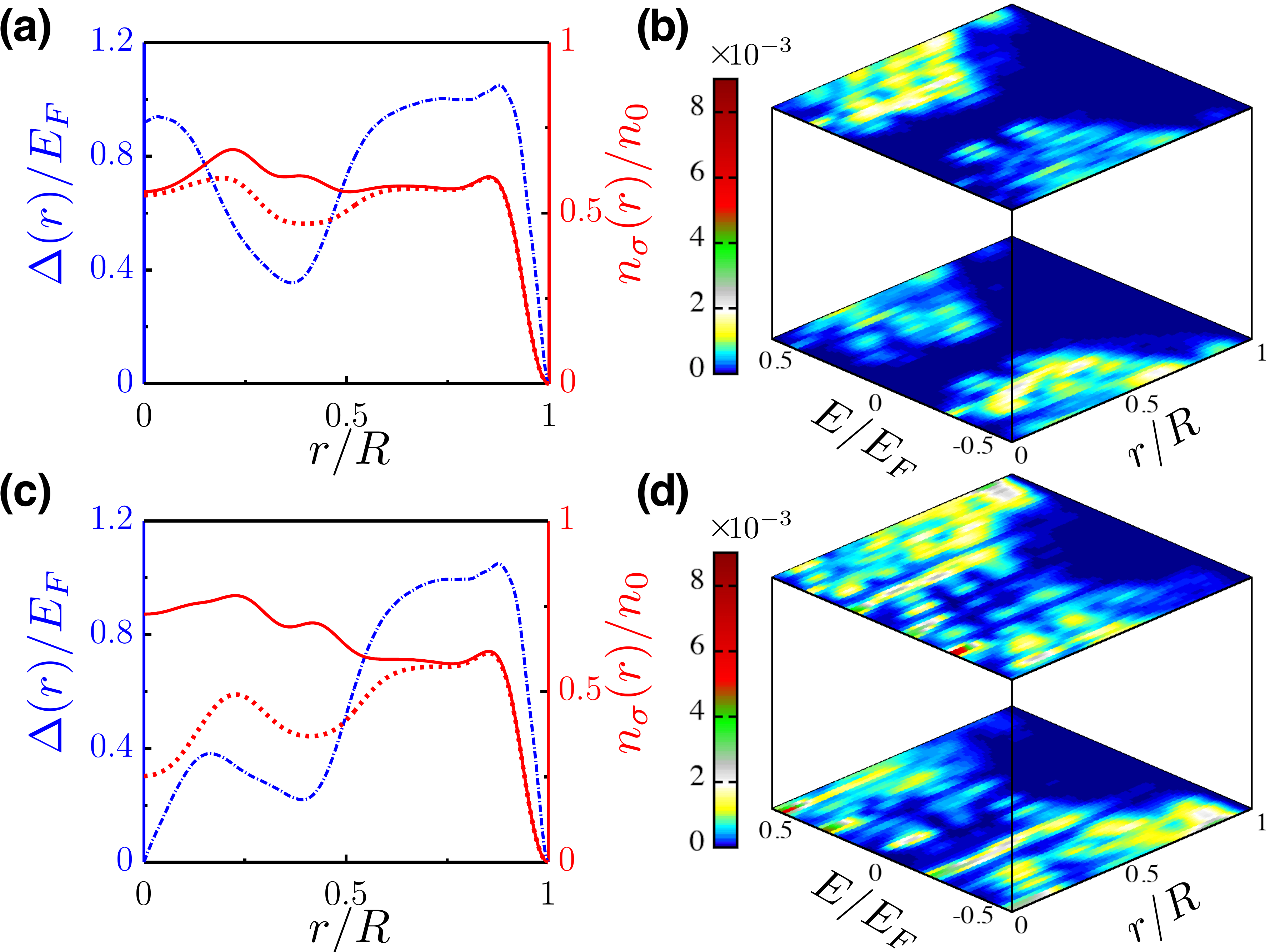}
\caption{Vortex core structure and local density of states $D_\sigma(r,E)$ under the Raman profile in Eq.~(\ref{eq:coreprof}), with $r_0/w=0.05$.
(a)(b) Ground state is a vortex-less superfluid, with $\delta/E_F=0.48$, $\Omega_0/E_F=1.2$. (c)(d) Ground state is a gapped vortex state, with
$\delta/E_F=0.84$ and $\Omega_0/E_F=1.2$. For all subplots, we take $k_Fw=15$, $k_F R=15$, with other parameters the same as those in Fig.~{\color{red}2} of the main text.}
\label{fig:smallcore}
\end{figure}

Given such a small core, we thus approximate the laser intensity using a Gaussian profile as discussed in the main text. To justify such an approximation, we explicitly calculate properties of a vortex state under the Raman coupling
\begin{eqnarray}
\Omega(r) =\left\{\begin{array}{@{}c}\Omega_0 (\frac{r}{r_0})^{2|l|} e^{-2\frac{r^2}{w^2}}, \quad \text{for}\,\,r<r_0\\
\Omega_0  e^{-2\frac{r^2}{w^2}}, \quad\quad \quad\quad\text{for}\,\, r\geqslant r_0\end{array},\right.\label{eq:coreprof}
\end{eqnarray}
where the size of the central core is given by $r_0$. In Fig.~\ref{fig:smallcore}, we show vortex core structures and local density of states for the case with $r_0/w=0.05$. Compared with Figs.~{\color{red}3} and {\color{red}5} in the main text, all key signatures of the vortex state are qualitatively the same. It is therefore justified to approximate the Raman coupling with a Gaussian function, as we have done in the main text.

\section{Effective single-particle Hamiltonian}

In this section, we briefly outline the derivation of the effective single-particle Hamiltonian ${\cal H}_s$ in the main text. We start from considering a pair of co-propagating Raman beams carrying different angular momenta, which couple two hyperfine states ($\uparrow$ and $\downarrow$) of an atom. After the adiabatic elimination of the excited states, the effective single-particle Hamiltonian in the hyperfine-spin basis reads
\begin{equation}
\mathcal{H}_{0}=
\left [\begin{array}{cc}-\frac{\hbar^2 \nabla^2}{2M}+V_{\text{ext}}({\bf r})+\frac{\delta}{2} & \Omega(r)e^{-i(l_1-l_2)\theta} \\ \Omega(r)e^{i(l_1-l_2)\theta} & -\frac{\hbar^2 \nabla^2}{2M}+V_{\text{ext}}({\bf r})-\frac{\delta}{2}
\end{array}\right],
\label{H0}
\end{equation}
where $\nabla^2=\frac{1}{r}\frac{\partial}{\partial r}\left(r\frac{\partial}{\partial r}\right)+\frac{1}{r^2}\frac{\partial ^2}{\partial \theta^2}$, and the phase windings $e^{-i l_i \theta}$ ($i=1,2$) reflect the orbital angular momenta
$-l_i \hbar$ carried by the two Raman beams. The two-photon Raman coupling and detuning are denoted by $\Omega(r)$ and $\delta$, respectively. The atoms are confined in an external potential denoted by
$V_{\text{ext}}({\bf r})$.

The phase terms in the off-diagonal components can be eliminated via a unitary transformation ${\cal H}_s=U^{\dag}{\cal H}_0 U$, where
$U={\rm diag}(e^{-il \theta},\  e^{il \theta})$, with $l=(l_1-l_2)/2$. We thus arrive at the effective single-particle Hamiltonian ${\cal H}_s$ given in the main text.

\begin{figure}
\includegraphics[width=0.8\columnwidth]{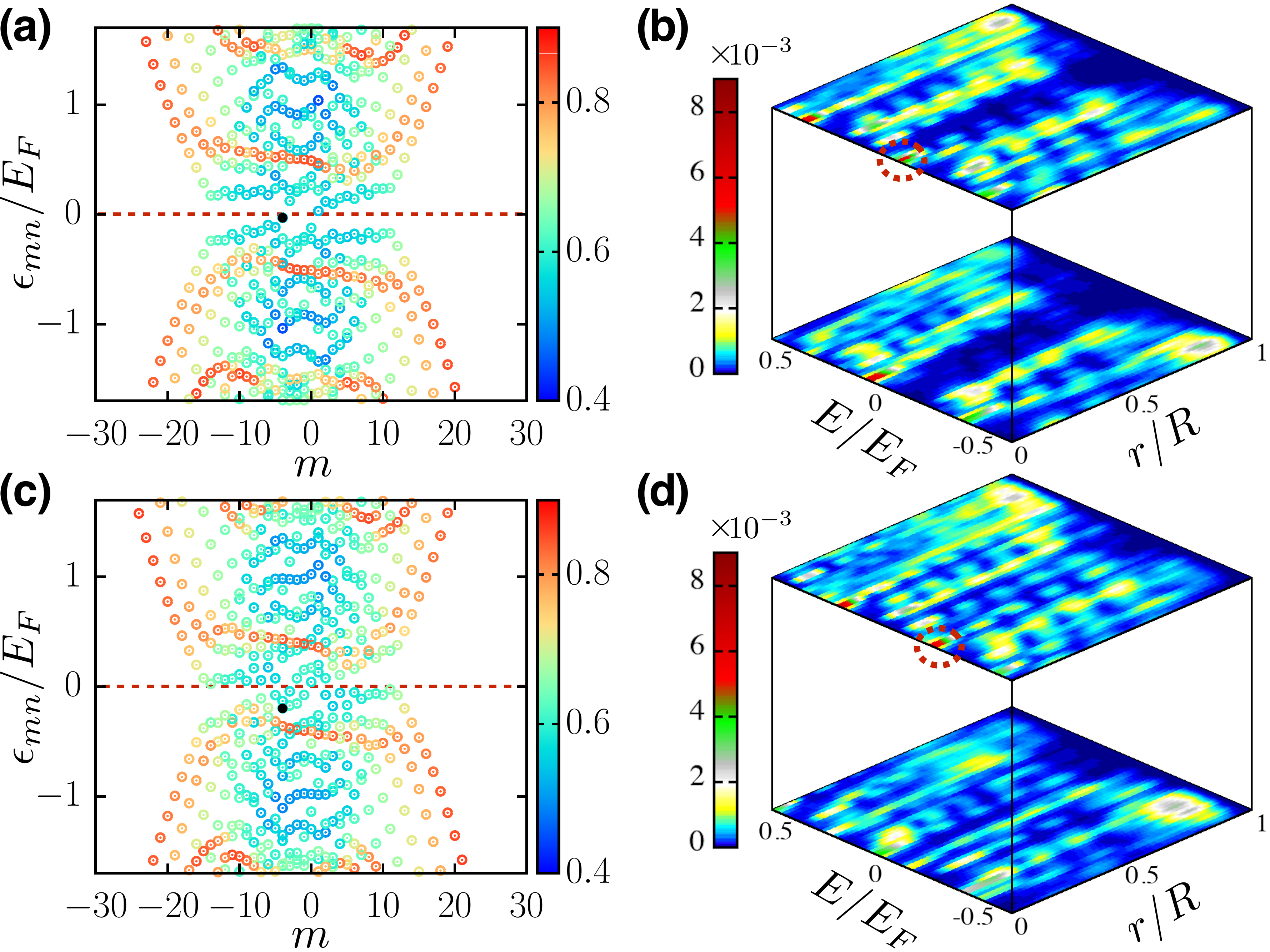}
\caption{Energy spectrum and local densities of state $D_{\sigma}(r, E)$ for (a)(b) gapped vortex state with $\delta/E_F=0.48$, and (c)(d) gapless vortex state with $\delta/E_F=0.84$.
For the fully-gapped vortex state in (a), most of the spin polarizations at the vortex core is carried by the occupied CdGM state marked in black, which is also circled out in $D_{\sigma}(r, E)$ (b).
For the gapless vortex in (c), spin polarization again concentrates near $r=0$ and is carried by a state (in black) that can be smoothly connected to the CdGM state in (a). In (b)(d), the upper (lower) layer shows the local densities of state of the spin-down (spin-up) component. Here we take $\Omega_0/E_F=2$, while other parameters are the same as those in Fig.~{\color{red}2}. See also Fig.~{\color{red}3} for the convention of color coding.}
\label{fig:S3}
\end{figure}

\section{Renormalizing the bare interaction}

The renormalization of the bare interaction strength $g$ is obtained by solving a two-body problem. In the absence of Raman beams, it is convenient to express the two-dimensional Hamiltonian in momentum space
\begin{eqnarray}
H & = & \sum_{{\bf k}\sigma} \varepsilon_{{\bf k}}^{\phantom{\dag}}a^{\dag}_{{\bf k}\sigma}a_{{\bf k}\sigma}^{\phantom{\dag}}-\frac{g}{S}\sum_{{\bf k}{\bf k}' {\bf q}}a^{\dag}_{{\bf k}+{\bf q},\uparrow}a^{\dag}_{{\bf k}'-{\bf q},\downarrow }a_{{\bf k}',\downarrow}^{\phantom{\dag}}a_{\bf k, \uparrow}^{\phantom{\dag}},
\end{eqnarray}
where $a_{{\bf k}\sigma}^{\phantom{\dag}}$ ($a^\dag_{{\bf k}\sigma}$) is the annihilation (creation) operator for a fermionic atom with momentum ${\bf k}$ and spin $\sigma$, $\varepsilon_{\bf k}^{\phantom{\dag}}=\hbar^2 {\bf k}^2 /(2M)$, and $S$ is the quantization area. The two-body bound state can be written as $| \Psi \rangle= \sum_{\bf k}\phi_{\bf k}a^{\dag}_{\bf k,\uparrow}a^{\dag}_{-\bf k,\downarrow} |\text{vac}\rangle$, where $\phi_{\bf k}$ is the bound-state wave function. Applying the Schr\"odinger's equation
$H |\Psi \rangle ={-E_B} |\Psi \rangle$ (with the two-body binding energy $E_B >0$), we obtain
\begin{eqnarray}
\frac{1}{g} & = & \frac{1}{S}\sum_{\bf k}\frac{1}{2\varepsilon_{\bf k}+E_B}.
\end{eqnarray}
The integral over the momentum ${\bf k}$ is divergent, which is a consequence of approximating the two-body interaction using a contract potential. We thus introduce a large-momentum cutoff $|{\bf k}|_c$ and obtain
$g = 4 \pi \hbar^2 /[M \ln (1+2 E_{c}/E_{B} )]$ as shown in the main text, with $E_c=\hbar^2 |{\bf k}|^2_{c}/(2M)$ being the cut-off energy. Since our model is renormalizable, all physical
results are independent of the choice of the high-energy cutoff $E_c$. The magnitude of $E_B$ thus reflects the interaction strength, with small (large) $E_B$ corresponding to weak (strong) interactions.
Note that while the renormalization relation is obtained in the absence of the synthetic gauge field, the same relation should hold under the spin-orbital-angular-momentum coupling, which is only linearly dependent on momentum~\cite{Cuirenorm}.

\begin{figure}
\includegraphics[width=0.8\columnwidth]{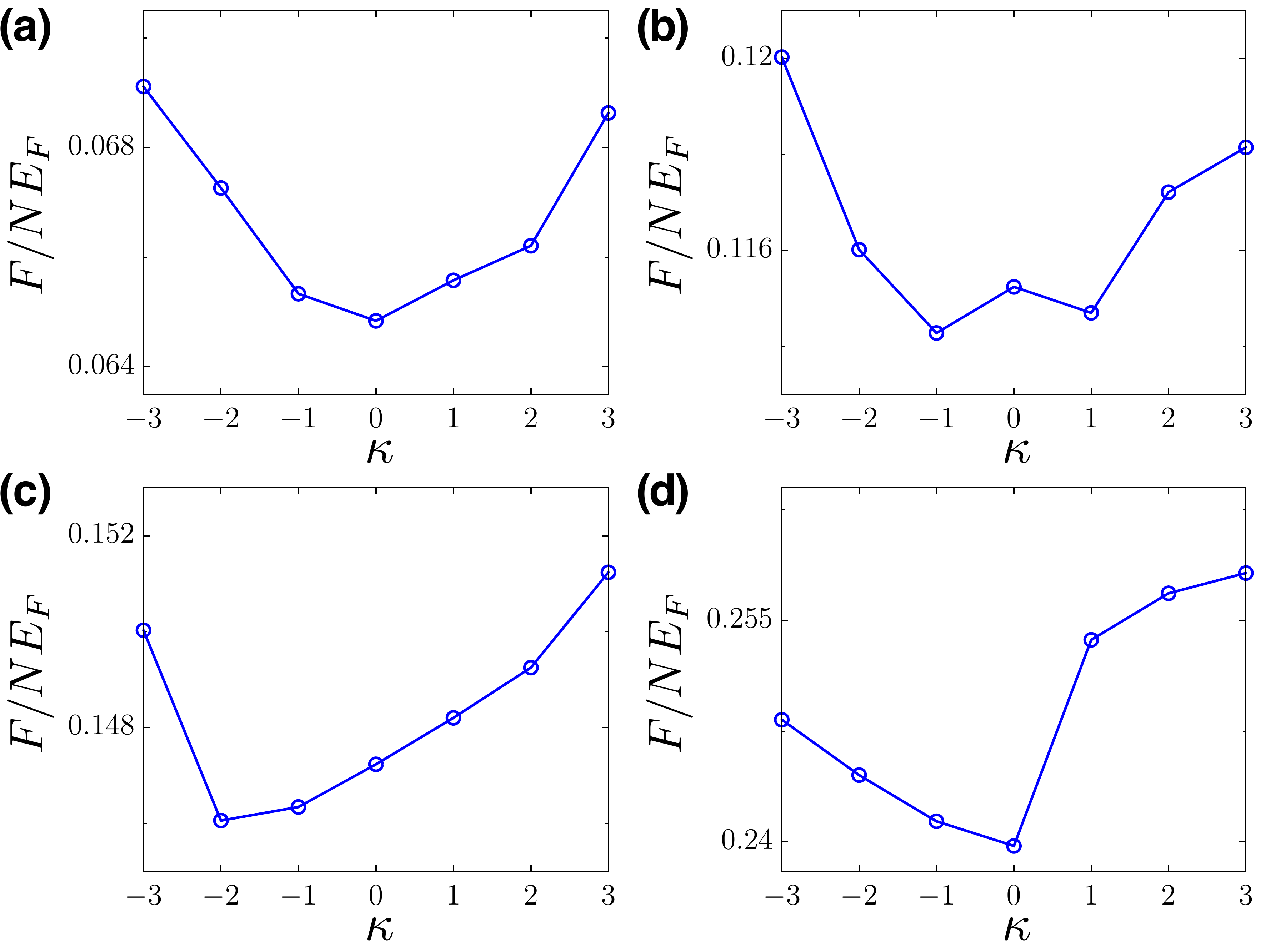}
\caption{Free energies of vortex states with different vorticity $\kappa$, under the parameters  $\delta/E_F=0.75$, $\Omega_0/E_F=2$, and with (a) $l=1$, (b) $l=3$, (c) $l=4$, (d) $l=7$. Other parameters are the same as those in Fig.~{\color{red}2} of the main text. }
\label{fig:S4}
\end{figure}

\section{BdG equation in the Bessel-function basis}

Based on the expansion of $u_{\sigma mn}$ and $v_{\sigma mn}$ in terms of the Bessel-function basis, as shown in the main text, the BdG equation becomes decoupled in each $m$ sector. For a given $m$, the BdG equation can be expressed as a matrix equation for the expansion coefficients $c^{(n')}_{\sigma mn}$ and $d^{(n')}_{\sigma mn}$,
\begin{eqnarray}\label{BDG1}
  \sum_{n''}
  \left[
    \begin{array}{cccc}
      K^{n'n''}_{\uparrow,m-l} & \Omega^{n'',m+l}_{n',m-l} & 0 & \Delta^{n'',m-l-\kappa}_{n',m-l} \\
      \Omega^{n'',m-l}_{n',m+l}& K^{n'n''}_{\downarrow,m+l} & -\Delta^{n'',m+l-\kappa}_{n',m+l}& 0 \\
      0 & -\Delta^{n'',m+l}_{n',m+l-\kappa} & -K^{n'n''}_{\uparrow,m+l-\kappa} & -\Omega^{n'',m-l-\kappa}_{n',m+l-\kappa} \\
      \Delta^{n'',m-l}_{n',m-l-\kappa} & 0 & -\Omega^{n'',m+l-\kappa}_{n',m-l-\kappa} & -K^{n'n''}_{\downarrow, m-l-\kappa} \\
    \end{array}
  \right]
  \left[
    \begin{array}{c}
      c^{(n'')}_{\uparrow mn} \\
      c^{(n'')}_{\downarrow mn} \\
      d^{(n'')}_{\uparrow mn} \\
      d^{(n'')}_{\downarrow mn} \\
    \end{array}
  \right]
  =\epsilon_{mn}
  \left[
    \begin{array}{c}
      c^{(n')}_{\uparrow mn} \\
      c^{(n')}_{\downarrow mn} \\
      d^{(n')}_{\uparrow mn} \\
      d^{(n')}_{\downarrow mn} \\
    \end{array}
  \right].
\end{eqnarray}
The elements are given by
\begin{align}
K^{n'n''}_{\sigma, p} & =  \Big(\frac{\hbar^2 \alpha^2_{n',p}}{2mR^2}+\tau\frac{\delta}{2} -\mu \Big)\delta_{n'n''}, \\
\Omega^{n'',q}_{n',p}&= \int r dr R_{n',p}(r)\Omega(r)R_{n'',q}(r),\\
\Delta^{n'',q}_{n',p}&=\int r dr R_{n',p}(r)\Delta(r)R_{n'',q}(r).
\end{align}
For a given profile of $\Delta(r)$, we diagonalize the matrix in Eq.~(\ref{BDG1}) to determine the quasiparticle spectrum $\epsilon_{mn}$ and the coefficients $c^{(n')}_{\sigma mn}$ and $d^{(n')}_{\sigma mn}$.  The profile
$\Delta(r)$ and the chemical potential $\mu$ are then determined through the self-consistent equation and the number equation given in the main text. The free energy of the pairing state can be evaluated as
\begin{align}
F=  \frac{1}{2}\sum_{mn}\epsilon_{mn}\left[\vartheta(-\epsilon_{mn})-\sum_\sigma\int d {\bf r}|v_{\sigma mn}(\bf r)|^2\right]
+\int d{\bf r}\frac{|\Delta({\bf r})|^2}{g}+\mu N.
\end{align}

\section{Energy spectrum and local density of states for gapless vortices}

For a fully-gapped vortex, the majority of the spin polarization at $r=0$ is carried by a single occupied ($\epsilon_{mn}<0$) CdGM state, as illustrated in Fig.~\ref{fig:S3}(a)(b). When increasing $\delta$, the vortex state can become gapless. Despite the closing of the bulk gap, the heavily-polarized CdGM state is smoothly connected to an extended state in the gapless spectrum, which is also heavily polarized at $r=0$ [see Fig.~\ref{fig:S3}(c)(d)]. As such, both gapped and gapless vortices are amenable to our proposed detection scheme.

\begin{figure}[tbp]
\includegraphics[width=0.8\columnwidth]{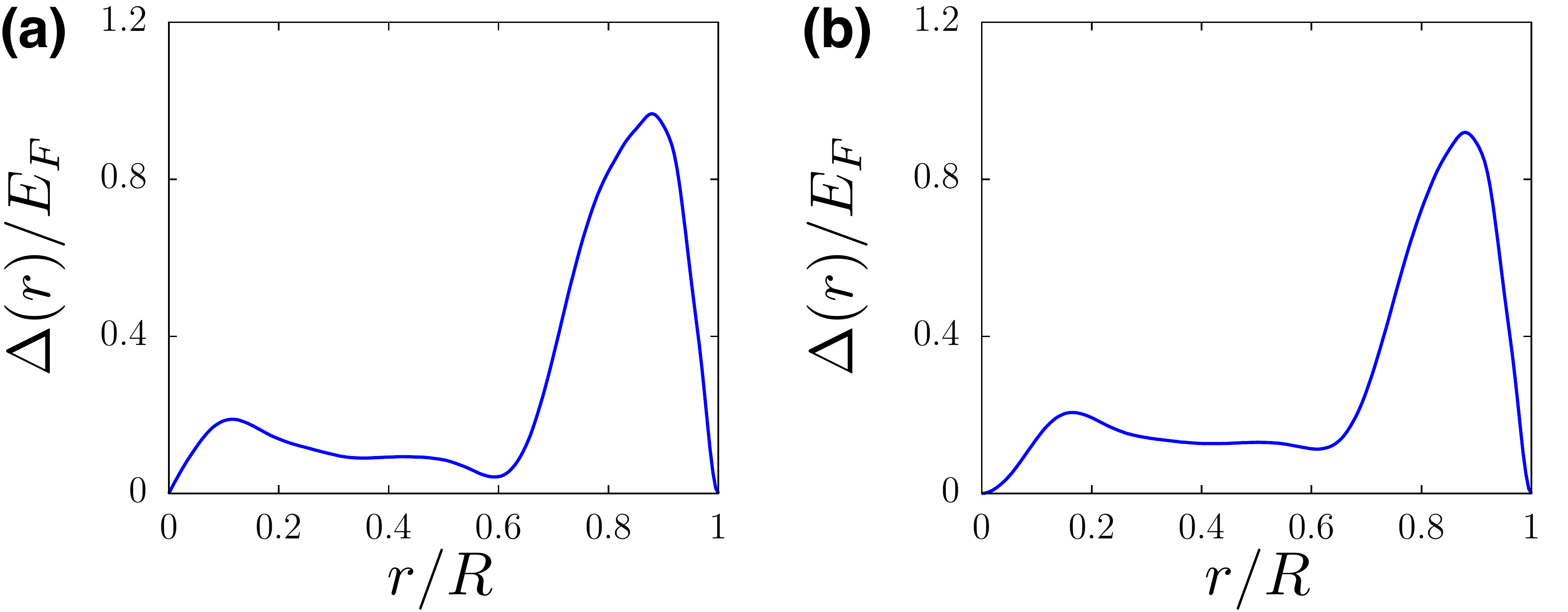}
\caption{Order parameter profiles for: (a) a vortex state with $\kappa=-1$, with $l=3$; and (b) a vortex state with $\kappa=-2$, with $l=4$. For both calculatoins, we take $\delta/E_F=0.75$ and $\Omega_0/E_F=2$. For both plots, we take $k_Fw=15$, $k_FR=15$, and other parameters are the same as those in Fig.~{\color{red}2} of the main text.}
\label{fig:S5}
\end{figure}

\section{Vorticity of vortex states}

In this section, we provide more details on the dependence of the vorticity $\kappa$ of the ground-state vortex on the angular-momentum transfer $2l\hbar$ of the spin-orbital-angular-momentum coupling. A typical calculation is shown in Fig.~\ref{fig:S4}, which indicates that the ground state is a vortex-less superfluid under small $l\hbar$ [as in Fig.~\ref{fig:S4}(a)]. With increasing $l\hbar$, the vorticity becomes finite [see Fig.~\ref{fig:S4}(b)(c)], and can even take values larger than unity [see Fig.~\ref{fig:S4}(c)]. For sufficiently large $l\hbar$, however, the ground state becomes vortex-less again [see Fig.~\ref{fig:S4}(d)].

These numerical results are consistent with the physical picture that the vortex formation
sensitively depends on the Fermi-surface deformation under the interplay of
spin-orbital-angular-momentum coupling (characterized by $l\hbar$) and Zeeman fields (characterized by $\Omega_0$ and $\delta$). Under fixed $\delta$ and $\Omega_0$, such a deformation is significant only for an intermediate $l\hbar$, but is small for either small or very large $l\hbar$. The situation is similar to that of spin-orbit-coupling induced Fulde-Ferrell pairing~\cite{socreview4}, with the key difference that, for spin-orbit-coupling induced Fulde-Ferrell states, Cooper pairs with a small center-of-mass momentum are stabilized in the presence of even an infinitesimally small Fermi-surface deformation (on the mean-field level). By contrast, since the angular momentum is quantized in our case, despite the small Fermi-surface deformation under small or large $l\hbar$ (under finite $\delta$), zero angular-momentum Cooper pairs are still stable, compared to those with finite $|\kappa|$. We further note that the case in Fig.~\ref{fig:S4}(c) raises the interesting question as to the stability of vortex state with $\kappa=-2$: whether such a vortex state would split up into two vortices each with $\kappa=-1$? We leave this interesting question to future studies.

Finally, in Fig.~\ref{fig:S5}, we compare order parameter profiles between vortex states with $\kappa=-1$ and $\kappa=-2$, with other parameters fixed. In both cases, vortex states are similar in size, indicating that the size of the vortex state is not sensitive to $\kappa$, but rather depends on the laser waist $w$ (see also Fig.~4 in the main text).


\end{document}